\documentclass[preprint2]{aastex}

\usepackage{graphicx}
\usepackage{natbib}

\newcommand{\thisgrb}{GRB\,051022}
\newcommand{\swift}{\textit{Swift}}
\newcommand{\hete}{HETE-2}
\newcommand{\swcommand}[1]{\textsf{#1}}
\newcommand{\obsname}[1]{#1}
\newcommand{\nh}{\ensuremath{N_\mathrm{H}}}
\newcommand{\plusminus}[2]{\ensuremath{^{+#1}_{-#2}}}
\newcommand{\ebv}{\ensuremath{E_{B-V}}}
\newcommand{\chisqred}{\ensuremath{\chi^2_{\mathrm{red}}}}
\newcommand{\chisqdof}{\ensuremath{\chi^2/\mathrm{d.o.f.}}}
\newcommand{\num}{\ensuremath{\nu_{\mathrm{m}}}}
\newcommand{\nuc}{\ensuremath{\nu_{\mathrm{c}}}}
\newcommand{\nua}{\ensuremath{\nu_{\mathrm{a}}}}
\newcommand{\Fp}{\ensuremath{F_{\nu,\mathrm{max}}}}
\newcommand{\tjet}{\ensuremath{t_{\mathrm{j}}}}
\newcommand{\Eiso}{\ensuremath{E_{\mathrm{iso}}}}
\newcommand{\thetajet}{\ensuremath{\theta_{\mathrm{j}}}}
\newcommand{\Epromptjet}{\ensuremath{E_{\mathrm{p,jet}}}}
\newcommand{\Epromptiso}{\ensuremath{E_{\mathrm{p,iso}}}}
\newcommand{\Ejet}{\ensuremath{E_{\mathrm{jet}}}}
\newcommand{\epse}{\ensuremath{\varepsilon_{\mathrm{e}}}}
\newcommand{\epsB}{\ensuremath{\varepsilon_{\mathrm{B}}}}
\newcommand{\Epeak}{\ensuremath{E_{\mathrm{peak}}}}

\shorttitle{\thisgrb\ as a prototype dark burst}
\shortauthors{Rol et al.}

\begin{document}

\title{\thisgrb: physical parameters and extinction of a prototype
  dark burst}

\author{
Evert~Rol\altaffilmark{1},
Alexander~van~der~Horst\altaffilmark{2},
Klaas~Wiersema\altaffilmark{2},
Sandeep~K.~Patel\altaffilmark{3,4},
Andrew~Levan\altaffilmark{5},
Melissa~Nysewander\altaffilmark{6},
Chryssa~Kouveliotou\altaffilmark{3,7},
Ralph~A.\,M.\,J.~Wijers\altaffilmark{2},
Nial~Tanvir\altaffilmark{1},
Dan~Reichart\altaffilmark{8},
Andrew~S.~Fruchter\altaffilmark{6},
John~Graham\altaffilmark{6,9},
Jan-Erik~Ovaldsen\altaffilmark{10},
Andreas~O.~Jaunsen\altaffilmark{10},
Peter~Jonker\altaffilmark{11,12,13},
Wilbert~van~Ham\altaffilmark{14},
Jens~Hjorth\altaffilmark{15},
Rhaana~L.\,C.~Starling\altaffilmark{1},
Paul~T.~O'Brien\altaffilmark{1},
Johan~Fynbo\altaffilmark{15},
David~N.~Burrows\altaffilmark{16},
Richard~Strom\altaffilmark{2,17}
}

\altaffiltext{1}{Department of Physics and Astronomy, University of
  Leicester, University Road, Leicester, LE1\,7RH, United Kingdom;
  evert.rol@star.le.ac.uk}

\altaffiltext{2}{Astronomical Institute, University of Amsterdam,
  Kruislaan 403, NL-1098~SJ Amsterdam, The Netherlands}

\altaffiltext{3}{National Space Science
  and Technology Center, 320 Sparkman Drive, Huntsville, AL-35805,
  USA}

\altaffiltext{4}{Optical Sciences Corporation, 6767 Old Madison Pike,
  Suite 650, Huntsville, AL, 35806}

\altaffiltext{5}{Department of Physics, University of Warwick,
  Coventry, CV4 7AL, UK}

\altaffiltext{6}{Space Telescope Science Institute, 3700 San Martin
  Drive, Baltimore, MD-21218}

\altaffiltext{7}{NASA Marshall Space Flight Center}

\altaffiltext{8}{Department of Physics and Astronomy, University of
  North Carolina at Chapel Hill, Campus Box 3255, Chapel Hill,
  NC-27599, USA}

\altaffiltext{9}{Department of Physics and Astronomy, Johns Hopkins
  University, 3400 North Charles St., Baltimore, MD-21218}

\altaffiltext{10}{Institute of Theoretical Astrophysics, University of
  Oslo, P.O.Box 1029, Blindern, N-0315 Oslo, Norway} 

\altaffiltext{11}{SRON, Netherlands Institute for Space Research,
  Sorbonnelaan 2, NL-3584~CA Utrecht, The Netherlands}

\altaffiltext{12}{Harvard--Smithsonian Center for Astrophysics, 60
  Garden Street, Cambridge, MA-02138, Massachusetts, USA}

\altaffiltext{13}{Astronomical Institute, Utrecht University, P.O.Box
  80000, 3508 TA, Utrecht, The Netherlands}

\altaffiltext{14}{Department of Astrophysics, Radboud University
  Nijmegen, P.O.Box 9010, NL-6500~GL Nijmegen, The Netherlands}

\altaffiltext{15}{Dark Cosmology Centre, Niels Bohr Institute,
  University of Copenhagen, Juliane Maries Vej 30, DK-2100 Copenhagen
  \O, Denmark}

\altaffiltext{16}{Penn State University, State College, PA 16801, USA}

\altaffiltext{17}{ASTRON, P.O. Box 2, NL-7990~AA Dwingeloo, Netherlands}

\begin{abstract}
  \thisgrb\ was undetected to deep limits in early optical
  observations, but precise astrometry from radio and X-ray showed
  that it most likely originated in a galaxy at $z\approx0.8$.  We
  report radio, optical, near infra-red and X-ray observations of
  \thisgrb.  Using the available X-ray and radio data, we model the
  afterglow and calculate the energetics of the afterglow, finding it
  to be an order of magnitude lower than that of the prompt emission.
  The broad-band modeling also allows us to precisely define various
  other physical parameters and the minimum required amount of
  extinction, to explain the absence of an optical afterglow. Our
  observations suggest a high extinction, at least $2.3$ magnitudes in
  the infrared ($J$) and at least 5.4 magnitudes in the optical ($U$) in
  the host-galaxy restframe. Such high extinctions are unusual for
  GRBs, and likely indicate a geometry where our line of sight to the
  burst passes through a dusty region in the host that is not directly
  co-located with the burst itself.

\end{abstract}


\section{Introduction} \label{section:intro}

Dark gamma-ray bursts (GRBs) --- at the most basic level those without
optical afterglows --- are a long-standing issue in GRB observations.
Although in many cases the non-detection of an afterglow at optical
wavelengths may simply be due to an insufficiently deep search, or one
which takes place at late times \citep[e.g.][]{fynbo2001:aa369:373}, a
subset of GRBs with bright X-ray afterglows remains undetected despite
prompt and deep optical searches \citep[e.g.][]{groot1998:apj493:27}
and directly implies suppression of the optical light.

There are several plausible explanations for this, the most likely
being that the burst is at high redshift, such that the Ly-alpha break
has crossed the passband in question, or that there is high extinction
in the direction of the GRB. Examples of both have been found, with a
small number of GRBs at $z>5$ appearing as $V$ and $R$ band dropouts
\citep[e.g.][]{jakobsson2006:aa447:897, haislip2006:nat440:181} and
some GRB afterglows appearing very red at lower redshift, due to
effects of extinction
\citep[e.g.][]{levan2006:apj647:471,rol2007:mnras374:1078}.

Identification of GRBs at very high redshifts is the key to using them
as cosmological probes.  The proportion of bursts exhibiting
high dust extinction is also interesting from the point of view of
estimating the proportion of star formation that
is dust enshrouded, as well as understanding the environments
which favor GRB production \citep{trentham2002:mnras334:983,mnras2004:352:1073}.

The detection and follow-up of dark bursts at other
wavelengths is essential, as it enables 1) the modeling of the
afterglow, deriving estimates of the extinction and energies involved,
potentially providing information about the direct burst environment, 2)
pinpointing the burst position in the host, to enable late-time
high resolution imaging and the detection of dust enhanced regions in
the host, and 3) determination of the properties of the GRB host itself,
such as the SFR and average host-galaxy extinction.

The High Energy Transient Explorer 2 mission (HETE-2;
\citealt{ricker2003:aipc662:3}) detected and located an unusually
bright gamma-ray burst \citep{olive2005:gcn4131}
with its three main instruments, the French Gamma Telescope (FREGATE),
the Wide field X-ray monitor (WXM) and the Soft X-ray Camera, (SXC), on
October 22, 2005. A 2.5 arcminute localization was sent out within
minutes, enabling prompt follow-up observations
\citep[e.g.][]{torii2005:gcn4130,schaefer2005:gcn4132}; a
target-of-opportunity observation was also performed with \swift.
Details of the HETE-2 observations can be found in
\citet{nakagawa2006:pasjl58:35}.

The \swift\ observations resulted in the detection of a single fading
point source inside the SXC error region, which was consequently
identified as the X-ray afterglow of \thisgrb\
\citep{racusin2005:gcn4141}. However, optical and near infra-red (nIR)
observations failed to reveal any afterglow to deep limits, while
radio and millimeter observations with the Very Large Array (VLA), the
Westerbork Synthesis Radio Telescope (WSRT) and the Plateau de Bure
Interferometer detected the radio counterpart
\citep{cameron2005:gcn4154,vanderhorst2005:gcn4158,bremer2005:gcn4157}.
The position coincides with its likely host galaxy
\citep{berger2005:gcn4148} at a redshift of $z = 0.8$
\citep{gal-yam2005:gcn4156}.

In this paper, we describe our X-ray, optical, nIR and radio
observations of \objectname[GRB 051022]{\thisgrb}.  The outline of the
paper is as follows: in Section \ref{section:obs} we describe our
observations, data reduction and initial results. In Section
\ref{section:analysis}, we analyze these results and form our
afterglow picture, which is discussed in Section
\ref{section:discussion}. Our findings are summarized in Section
\ref{section:conclusions}.

In the following, we have used $F \propto \nu^{-\beta} t^{-\alpha}$ in
our definition of $\alpha$ and $\beta$. We assume a cosmology with
$H_0 = 71\, \mathrm{km} \mathrm{s}^{-1} \mathrm{Mpc}^{-1}$, $\Omega_M =
0.27$ and $\Omega_{\Lambda} = 0.73$. All quoted errors in this paper
are 1 sigma (68\%) errors.

\section{Observations and data reduction} \label{section:obs}

\subsection{X-ray observations} \label{section:obs:X-ray}

X-ray observations were performed with the \swift\ X-Ray Telescope
(XRT) and the Chandra X-ray Observatory (CXO).

The XRT started observing the afterglow of \thisgrb\ 3.46 hours after
the \hete\ trigger, for a total effective integration time of 137~ks
between October 22 and November 6.

Observations were performed in Photon Counting (PC) mode, the most
sensitive observing mode. We reduced the data using the \swift\
software version 2.6 in the HEAsoft package version 6.2.0. Data were
obtained from the quick-look site and processed from level 1 to level
2 FITS files using the \swcommand{xrtpipeline} tool in its standard
configuration.  The first two orbits (until $2.1 \times 10^4$ seconds
post burst) show pile-up and were therefore extracted with an annular
rather than circular region, with an inner radius of 19 and 12\arcsec\
for orbits 1 and 2, respectively, and an outer radius of 71\arcsec.
Orbits 3~--~7 ($2.4 \times 10^4$~--~$4.9 \times 10^4$ seconds) were
extracted with a circular region of 71\arcsec\ radius, and later
orbits were extracted using a 47\arcsec\ radius circle instead. The
data for the light curve were extracted between channels 100 and 1000,
corresponding to 1 and 10 keV, respectively; while the commonly used
range is 0.3~--~10 keV, the large absorption prevents the detection of
any data from the source below 1 keV.  Otherwise, the procedure is
similar to that described in \citet{evans2007:aap469:379}.

Observations with the CXO started on October 25, 2005, 21:14:20, 3.34
days after the HETE trigger, for a total integration time of 20~ks
\citep{patel2005:gcn4163}.  Data were reduced in a standard fashion
with the CIAO package.

We performed astrometry by matching X-ray sources with an optical
$R$-band image that was astrometrically calibrated to the 2MASS
catalog. Our CXO position is RA, Dec = 23:56:04.115, +19:36:24.04
(J2000), with positional errors of 0.33\arcsec\ and 0.12\arcsec\ for
the Right Ascension and Declination, respectively. This puts the
afterglow within 0.5\arcsec\ of the center of its host galaxy.

We modeled the XRT spectra with an absorbed power law in XSpec
\citep{arnaud1996:proc:17}, using data from the first seven orbits. A
good fit ($\chisqdof = 87.2/99$) was obtained with a resulting
spectral energy index of $\beta = 1.00 \pm 0.12$ and excess absorption
(at $z = 0.8$ and for assumed Galactic abundances) of $\nh = (2.82 \pm
0.46) \times 10^{22}\ \mathrm{cm}^{-2}$ on top of the estimated
Galactic absorption at this position ($\nh = 4.06 \times 10^{20}\
\mathrm{cm}^{-2}$, \citealt{dickey1990:araa28:215}). The CXO data are
fully in agreement with these values, showing no change in the
spectrum over time between 0.3 and 3.3 days after the burst. The
absorption measured is far less than that measured by the HETE team in
their prompt data, $\nh = (8.8\plusminus{1.9}{1.8}) \times
10^{22}\,\mathrm{cm}^{-2}$ \citep{nakagawa2006:pasjl58:35}. This could
indicate a change in absorption between the early (prompt)
measurements and those at the time of the XRT observations. For the
prompt emission spectrum, however, the values found by Konus-Wind
\citep{golenetskii2005:gcn4150} are rather different than those found
by HETE-2, and may be the result of the lower energy cut-off for
FREGATE compared to Konus-wind. Alternatively, the fact that these
spectra are an average over the whole emission period may also result
in incorrect model parameters. In the two last cases, the \nh\ in the
prompt emission could be as low as the XRT value and still produce an
equally well fit, but with slightly different model parameters.

For the XRT data, \citet{butler2005:gcn4165} and
\citet{nakagawa2006:pasjl58:35} find a value somewhat higher than our
value ($4.9 \times 10^{22}\,\mathrm{cm}^{-2}$ and $5.3 \times
10^{22}\,\mathrm{cm}^{-2}$ respectively, when scaled by $(1+z)^3$,
\citealt{gunn1965:apj142:1633}). This difference could be explained by
a different count-binning or an updated XRT calibration used in our
modeling.

The XRT light curve count rates have been converted to 1--10~keV
fluxes using the results from our spectral modeling and calculating
the ratio of the flux and count rate at the logarithmic center of the
orbits. The 1~--~10 keV CXO flux was derived using the actual spectral fit.

A broken power law fit to the X-ray light curve results in $\alpha_1 =
1.16 \pm 0.06$, $\alpha_2 = 2.14 \pm 0.17$ and a break time of
$110\plusminus{21}{23}$ ks, or around 1.27 days. The difference between
$\alpha_1$ and $\alpha_2$, and the fact that the spectral slope does
not change across the break (the CXO measurement is past the break),
are highly indicative that the observed break in the light curve is a jet
break. In Section \ref{section:broadband-modeling}, we perform full
modeling of the afterglow using the fireball model, indeed resulting
in a jet-break time \tjet\ that agrees reasonably well with the break time as
determined from only the X-rays.  We point out that our value for
\tjet\ is different than that cited in \citet{racusin2005:gcn4169},
largely because their measurement of \tjet\ was based on a preliminary
XRT light curve.

\subsection{Optical and near infra-red observations} 
\label{section:obs:optical}

Observations were obtained in $Z$ and $R$-band with the William
Herschel Telescope (WHT) using the Auxiliary Port and the Prime Focus
Imaging Camera, respectively, in $r^{\prime}i^{\prime}z^{\prime}$ with
the Gemini South telescope using the GMOS instrument, in $JHK_s$ with
the Wide Field Camera on the United Kingdom InfraRed Telescope
(UKIRT), in $BVRI$ with the DFOSC instrument on the Danish 1.54m
telescope and in $J$ and $K_s$ with the Southern Astrophysical
Research (SOAR) telescope using OSIRIS. The optical data were reduced
in a standard fashion using the \swcommand{ccdproc} package within the
IRAF software \citep{tody1986:proc:733}, whereas the SOAR data were
reduced using the \swcommand{cirred} package within IRAF. The UKIRT
data were reduced using the standard pipeline reduction for WFCAM.

Photometric calibration was done using the calibration provided by
\citet{henden2005:gcn4184} for Johnson-Cousins filters. For the
$r^{\prime}i^{\prime}z^{\prime}$ GMOS filters, we converted the
magnitudes of the calibration stars provided by Henden to the Sloan
filter system using the transformations provided by
\citet{jester2005:aj130:873}, and verified by the published GMOS zero
points. The WHT $Z$-band was calibrated using the spectroscopic
standard star SP2323+157.  Calibration of the infrared $JHK$
magnitudes was done using the 2MASS catalog
\citep{skrutskie2006:aj131:1163}.

No variable optical source was found at the position of the X-ray and
radio afterglow. For the early epoch images ($<1$ day post burst), we
estimated a limiting magnitude by performing image subtraction between
this and a later image using the ISIS image subtraction package
\citep{alard2000:aaps144}. To this end, artificial low signal-to-noise
sources were added onto the images, with a Gaussian PSF matched in
size to the seeing (some artificial sources were added on top of
existing sources, e.g. galaxies, some on the background sky). We
determined our upper limit to be the point where we could retrieve
50\% of the artificial sources in the subtracted image. This assumes that
the change in brightness of any point source on top of the host
galaxy is sufficient to be seen in such a subtracted image. With the
difference in time between the epochs, this seems a reasonable
assumption (for example, for a source fading with a shallow power law like
slope of $F \propto t^{-0.5}$, the magnitude difference between the
two WHT $Z$-band observations is $\approx 0.6$ magnitudes).

Photometry of the host galaxy has been performed using aperture
photometry, with an aperture 1.5 times the seeing for each image,
estimated from the measured FWHM of the PSF for point sources in the
images.

Table \ref{table:obslog} shows the log of our optical/nIR
observations, while Table \ref{table:limmags} shows the upper limits
for any optical/nIR afterglow.

\begin{deluxetable}{lccccc}
  \tablecaption{Overview of optical observations\label{table:obslog}}
\tablehead{
\colhead{Start date} &
\colhead{$\Delta T$ (average)} &
\colhead{exposure time} &
\colhead{filter} &
\colhead{seeing} &
\colhead{telescope \& instrument} \\
&
\colhead{(days)} &
\colhead{(seconds)} &
&
\colhead{(arcsec)} &
}
\startdata
2005-10-22T23:25:14  &  0.4287   &  1800  &  $Z$      &  0.8  &  WHT + API        \\
                                                      
2005-10-23T00:22:33  &  0.4684   &  1620  &  $J$      &  1.2  &  SOAR + OSIRIS    \\
2005-10-23T00:56:00  &  0.4917   &  1620  &  $K_s$    &  1.3  &  SOAR + OSIRIS    \\
                                                      
2005-10-23T00:48:03  &  0.5144   &  1920  &  $i^{\prime}$     &  0.6  &  Gemini South + GMOS   \\
2005-10-23T01:07:53  &  0.5288   &  1920  &  $r^{\prime}$     &  0.6  &  Gemini South + GMOS   \\
2005-10-23T01:27:46  &  0.5426   &  1920  &  $z^{\prime}$     &  0.5  &  Gemini South + GMOS   \\
                                                      
2005-10-23T06:31:03  &  0.7525   &  720   &  $J$      &  1.4  &  UKIRT + WFCAM    \\
2005-10-23T06:36:39  &  0.7526   &  360   &  $H$      &  1.3  &  UKIRT + WFCAM    \\
2005-10-23T06:47:59  &  0.7604   &  360   &  $K$      &  1.3  &  UKIRT + WFCAM    \\
                                                      
2005-10-23T21:15:57  &  1.3389   &  1200  &  $Z$      &  1.0  &  WHT + API        \\
                                                      
2005-10-24T09:35:10  &  1.8467   &  720   &  $K$      &  0.3  &  UKIRT + WFCAM    \\
                                                      
2005-10-25T01:34:03  &  2.5181   &  1602  &  $K_s$    &  1.3  &  SOAR + OSIRIS    \\
2005-10-25T02:13:18  &  2.5454   &   720  &  $J$      &  1.2  &  SOAR + OSIRIS    \\
                                                      
2005-10-25T02:22:02  &  2.5698   &  1920  &  $r^{\prime}$     &  1.1  &  Gemini South + GMOS   \\
2005-10-25T02:39:59  &  2.5792   &  1440  &  $z^{\prime}$     &  1.2  &  Gemini South + GMOS   \\
                                                      
2005-10-26T00:36:58  &  3.4785   &  1800  &  $R$      &  1.4  &  WHT+PFIP         \\

2005-10-26T02:48:06  &  3.5695   &   600  &  Gunn $i$ &  1.4  &  DK1.54m + DFOSC  \\
2005-10-26T03:23:35  &  3.5942   &   600  &  $R$      &  1.9  &  DK1.54m + DFOSC  \\
2005-10-27T01:01:04  &  4.4952   &   600  &  $B$      &  2.3  &  DK1.54m + DFOSC  \\
2005-10-27T02:59:20  &  4.5773   &   600  &  $R$      &  1.6  &  DK1.54m + DFOSC  \\
2005-10-27T02:00:48  &  4.5367   &   600  &  $V$      &  1.8  &  DK1.54m + DFOSC  \\
2005-10-28T02:18:38  &  5.5491   &   600  &  $i$      &  1.4  &  DK1.54m + DFOSC  \\
2005-10-30T02:32:59  &  7.5590   &   600  &  $B$      &  1.8  &  DK1.54m + DFOSC  \\
2005-10-30T04:18:30  &  7.6323   &   600  &  $U$      &  1.8  &  DK1.54m + DFOSC  \\
2005-10-30T01:33:57  &  7.5180   &   600  &  $V$      &  1.4  &  DK1.54m + DFOSC  \\
2005-10-31T03:19:05  &  8.5910   &   600  &  $B$      &  1.0  &  DK1.54m + DFOSC  \\
2005-10-31T01:03:40  &  8.4970   &   600  &  $R$      &  1.0  &  DK1.54m + DFOSC  \\
2005-10-31T02:10:02  &  8.5431   &   600  &  $V$      &  1.0  &  DK1.54m + DFOSC  \\
2005-11-01T01:52:57  &  9.5312   &   600  &  $R$      &  0.9  &  DK1.54m + DFOSC  \\
2005-11-02T02:04:47  &  10.539   &   600  &  $V$      &  1.2  &  DK1.54m + DFOSC  \\
2005-11-03T01:10:34  &  11.502   &   600  &  $B$      &  1.2  &  DK1.54m + DFOSC  \\
2005-11-07T01:25:30  &  15.512   &   600  &  Gunn $i$      &  1.4  &  DK1.54m + DFOSC  \\
2005-11-08T01:40:48  &  16.523   &   600  &  Gunn $i$      &  1.4  &  DK1.54m + DFOSC  \\
\enddata
\end{deluxetable}

\begin{deluxetable}{ccccc}
\tablecaption{Limiting magnitudes\label{table:limmags}}
\tablehead{
\colhead{filter} &
\colhead{limiting magnitude\tablenotemark{a}} &
\colhead{$\Delta T$ (average)} &
\colhead{frequency} &
\colhead{specific flux\tablenotemark{b}} \\
&
&
\colhead{days} &
\colhead{Hz} &
\colhead{$\mu$Jy}
}
\startdata
$K_s$    &  $>20.0$   &  0.4917    &    $1.40 \cdot 10^{14}$   &   $< 6.82 $  \\
$J$      &  $>20.3$   &  0.4684    &    $2.40 \cdot 10^{14}$   &   $<12.3  $  \\
$Z$      &  $>22.9$   &  0.4287    &    $3.43 \cdot 10^{14}$   &   $< 2.66 $  \\
$z^{\prime}$     &  $>23.5$   &  0.5426    &    $3.36 \cdot 10^{14}$   &   $< 1.53 $  \\
$r^{\prime}$     &  $>25.3$   &  0.5288    &    $4.76 \cdot 10^{14}$   &   $< 0.305$  \\
\enddata
\tablenotetext{a}{See text for the definition of the limiting
  magnitude.}  \tablenotetext{b}{Specifc fluxes have been corrected for a
  Galactic extinction value of $E_{B-V} = 0.04$
  \citep{schlegel1998:apj500:525}, and converted from magnitudes using
  the calibration by \citet{tokunaga2005:pasp117:421} for the $JK_s$
  filters; the other filters are on the magnitude AB-system
  \citep{oke1983:apj226:713}}
\end{deluxetable}

\subsection{Radio observations} \label{section:obs:radio}

Radio observations were performed with the WSRT at 8.4 GHz, 4.9 GHz
and 1.4 GHz.  We used the Multi Frequency Front Ends
\citep{tan1991:aspc19:42} in combination with the IVC+DZB back
end\footnote{See sect. 5.2 at
  \url{http://www.astron.nl/wsrt/wsrtGuide/node6.html}} in continuum mode,
with a bandwidth of 8x20 MHz. Gain and phase calibrations were
performed with the calibrators 3C~286 and 3C~48, although at one 8.4
GHz measurement 3C~147 was used. Reduction and analysis were performed
using the MIRIAD software
package\footnote{\url{http://www.atnf.csiro.au/computing/software/miriad}}.
The observations are detailed in Table \ref{table:radiolog}. In our
modeling described in section \ref{section:broadband-modeling} we
have also used the VLA radio detection at 8.5 GHz from
\citet{cameron2005:gcn4154}.

\begin{deluxetable}{lcccc}
  \tablecaption{Overview of \obsname{WSRT} radio
    observations\label{table:radiolog}}
\tablehead{
\colhead{Start date} &
\colhead{$\Delta T$ (average)} &
\colhead{integration time} &
\colhead{frequency} &
\colhead{specific flux} \\
&
\colhead{(days)} &
\colhead{(hours)} &
\colhead{(GHz)} &
\colhead{($\mu$Jy)}
}

\startdata
2005-11-04T18:14:24  &  13.37   &  4.0   &  8.5   &  38 $\pm$ 132 \\
2005-11-08T14:19:41  &  17.19   &  7.0   &  8.5   &  28 $\pm$  97 \\
2005-10-23T15:20:10  &   1.19   &  5.0   &  4.9   & 281 $\pm$  32 \\
2005-10-24T15:17:17  &   2.22   &  6.2   &  4.9   & 342 $\pm$  34 \\
2005-10-25T15:12:58  &   3.30   &  5.4   &  4.9   & 143 $\pm$  30 \\
2005-10-28T18:33:08  &   6.40   &  8.5   &  4.9   &  91 $\pm$  28 \\
2005-10-30T18:00:00  &   8.32   &  5.8   &  4.9   & 138 $\pm$  28 \\
2005-11-01T18:00:00  &  10.38   &  8.9   &  4.9   & 169 $\pm$  28 \\
2005-11-04T17:31:12  &  13.37   &  4.6   &  4.9   &  70 $\pm$  34 \\
2005-10-25T15:56:10  &  3.33    &  5.4   &  1.4   &   8 $\pm$  78 \\
\enddata
\end{deluxetable}

\section{Analysis} \label{section:analysis}

\subsection{Broadband modeling}
\label{section:broadband-modeling}

We have performed broadband modeling of the X-ray and radio
measurements, using the methods presented in \citet{vanderhorst2007:submitted}.  In our modeling we assume a purely
synchrotron radiation mechanism.

The relativistic blastwave causing the afterglow accelerates electrons
to relativistic velocities, which gives rise to a broadband spectrum
with three characteristic frequencies: the peak frequency \num,
corresponding to the minimum energy of the relativistic electrons that
are accelerated by the blastwave, the cooling frequency \nuc,
corresponding to the electron energy at which electrons lose a
significant fraction of their energy by radiation on a timescale that
is smaller than the dynamical timescale, and the self-absorption
frequency \nua, below which synchrotron self-absorption produces
significant attenuation.  The broadband spectrum is further
characterized by the specific peak flux \Fp\ and the slope $p$ of the electron
energy distribution.

The dynamics of the relativistic blastwave determine the temporal
behavior of the broadband synchrotron spectrum, i.e. the light curves
at given frequencies.  At first the blastwave is extremely
relativistic, but is decelerated by the surrounding medium.  When the
Lorentz factor $\Gamma$ of the blastwave becomes comparable to
$\thetajet^{\,-1}$, where \thetajet\ is the opening angle of the jet, the
jet starts to spread sideways. At that time, \tjet, the temporal
behavior of the broadband spectrum changes \citep[see
e.g.][]{rhoads1997:apj487:487}.

We fit our data to six parameters: \nuc, \num, \nua, \Fp, $p$ and
\tjet.  From these parameters and the redshift of the burst, $z =
0.8$, we can find the physical parameters governing the blastwave and
its surroundings: the blastwave isotropic equivalent energy \Eiso, the
jet opening angle \thetajet, the collimation corrected blastwave
energy \Ejet, the fractional energy densities behind the relativistic
shock in electrons and in the magnetic field, \epse\ and \epsB\
respectively, and the density of the surrounding medium. The meaning
of the latter parameter depends on the density profile of the surrounding
medium. For a homogeneous circumburst medium, we simply determine the density
$n$. For a massive stellar wind, where the density is proportional to
$R^{\,-2}$ with $R$ the distance to the GRB explosion center, we
obtain the parameter $A_{\ast}$, which is the ratio of the mass-loss
rate over the terminal wind velocity of the GRB progenitor.

\begin{figure*}
\includegraphics[width=\textwidth]{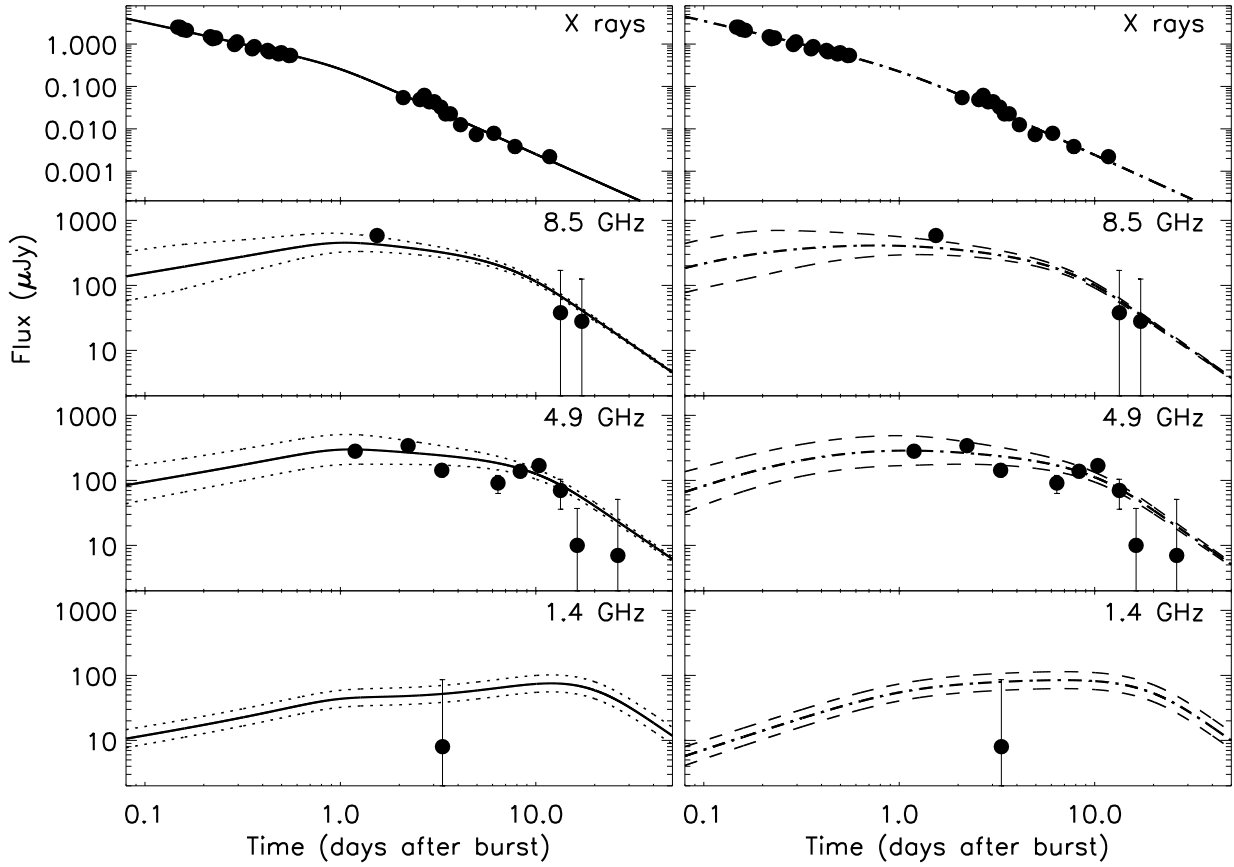}
\caption{Fit results for a homogeneous circumburst medium (left panel)
  and a massive stellar wind (right panel).  The solid and dash-dotted
  lines are the best model fits, and the dotted and dashed lines
  indicate the predicted rms scatter due to interstellar
  scintillation; see the appendix for further details. Also included in the figure (and modeling) is the
  reported VLA 8.5~GHz detection
  \citep[][left-most point in the 8.5~GHz subplot]{cameron2005:gcn4154}. \label{figure:resultsall}}
\end{figure*}

Our modeling results are shown in Table \ref{table:fitresults}, for
both the homogeneous external medium and the stellar wind environment.
The light curves for the best fit parameters are shown in Figure
\ref{figure:resultsall}.  We have performed Monte Carlo simulations
with synthetic data sets in order to derive accuracy estimates of the
best fit parameters, which are also given in the table.  It is evident
from the results that our six fit parameters are reasonably well
constrained in both cases for the circumburst medium.  The derived
physical parameters are also well constrained, except for \epse\ and
\epsB. The values we find for both the isotropic and the collimation
corrected energy, are similar to those found for other bursts; this is
also true for $p$. See e.g. \citet{panaitescu2001:apj560:49} and
\citet{yost2003:apj597:459}.  The jet opening angle and the density of
the surrounding medium are quite small, but both not unprecedented.
The jet-break time \tjet\ is somewhat smaller than estimated in
Section \ref{section:obs:X-ray}, but both estimates have relatively
large errors, likely because of the lack of (X-ray) data around the
jet-break time.

With the absence of optical light curves, it is not possible to
discriminate between the two different circumburst media.  This is
mainly due to the fact that the X-ray band lies above both \num\ and
\nuc, in which case the slopes of the light curves do not depend on
the density profile of the circumburst medium (even at 0.15 days,
back-extrapolating \nuc\ from Table \ref{table:fitresults} results in
its value being below the X-ray band). The \chisqred\ is somewhat
better for the stellar wind case, but the homogeneous case cannot be
excluded.  From the X-ray light curve, however, one can conclude that
the density profile of the medium does not change between
approximately 0.15 and 12 days after the burst.  If there were a
transition from a stellar wind to a homogeneous medium, the X-ray flux
has to rise or drop significantly, unless the densities are the
fine-tuned at the transition point \citep{peer2006:apj643:1036}.  From
the fact that the medium does not change during the X-ray
observations, one can draw conclusions on the distance of the wind
termination shock of the massive star: if one assumes that the medium
is already homogeneous at $\approx\,0.15$ days, the wind termination
shock position is at $R_{\mathrm{w}}\lesssim 9.8\cdot
10^{17}\,\mathrm{cm}$ (0.32 pc); if the circumburst medium is a
stellar wind up to $\approx 12$ days after the burst, $R_{\mathrm{w}}
\gtrsim 1.1\cdot 10^{19}\,\mathrm{cm}$ (3.7 pc).

\begin{deluxetable}{lll}
  \tablecaption{Results of broadband modeling for both a homogeneous
    external medium and a massive stellar wind.  The best fit
    parameters are shown together with accuracy estimates from Monte
    Carlo simulations with synthetic data sets.  The characteristic
    frequencies of the synchrotron spectrum and the specific peak flux are
    given at $t_{\mathrm{j}}$.  
\label{table:fitresults}
}
\tablehead{
\colhead{Parameter} &
\colhead{Homogeneous} &
\colhead{Stellar wind}
}
\startdata
\nuc(\tjet) & 
$(1.45\plusminus{1.12}{0.23}) \cdot 10^{17}$ Hz                  
& $(2.84\plusminus{0.32}{1.30}) \cdot 10^{17}$ Hz \\

\num(\tjet)     
& $(3.50\plusminus{2.26}{1.47}) \cdot 10^{11}$ Hz                  
& $(2.90\plusminus{2.03}{1.15}) \cdot 10^{11}$ Hz \\

\nua(\tjet)     
& $(4.56\plusminus{2.85}{3.08}) \cdot 10^{9}$ Hz                   
& $(2.68\plusminus{2.17}{1.60}) \cdot 10^{9}$ Hz \\

\Fp(\tjet)      
& $888\plusminus{52}{109}$ $\mu$Jy                                 
& $694\plusminus{30}{240}$ $\mu$Jy \\ 

$p$             
& $2.06\plusminus{0.19}{0.05}$                                     
& $2.10\plusminus{0.08}{0.09}$ \\

\tjet           
& $0.96\plusminus{0.40}{0.28}$ days                                
&$1.06\plusminus{0.41}{0.11}$ days \\

\thetajet       
& $3.39\plusminus{2.02}{2.27}$ deg                                
& $2.30\plusminus{1.09}{0.85}$ deg \\

\Eiso           
& $(5.23\plusminus{1.13}{1.69}) \cdot 10^{52}$ erg                 
& $(28.2\plusminus{31.0}{10.4}) \cdot 10^{52}$ erg \\

\Ejet         
& $(0.917\plusminus{0.655}{0.512}) \cdot 10^{50}$ erg                 
& $(2.27\plusminus{2.25}{0.79}) \cdot 10^{50}$ erg \\

\epse           
& $0.247\plusminus{1.396}{0.212}$                                  
& $0.0681\plusminus{0.3951}{0.0348}$ \\

\epsB           
& $(7.63\plusminus{42.57}{6.30}) \cdot 10^{-3}$                    
& $(8.02\plusminus{28.18}{7.17}) \cdot 10^{-3}$ \\

$n$             
& $(1.06\plusminus{9.47}{1.04}) \cdot 10^{-2}$ $\mathrm{cm}^{-3}$  
&  \nodata  \\

$A_{\ast}$$^a$  
& \nodata                                                    
& $(2.94\plusminus{6.98}{2.11}) \cdot 10^{-2}$  \\

\chisqred       
& $1.9$ 
& $1.5$ \\

\enddata
\tablenotetext{a}{The parameter $A_{\ast}$ is a measure
    for the density in the case of a stellar wind environment, being
    the ratio of the mass-loss rate over the terminal wind velocity,
    and here given in units of $10^{-5}$ Solar masses per year divided
    by a wind velocity of 1000 km/s \citep[see][]{vanderhorst2007:submitted}. }
\end{deluxetable}

\placefigure{figure:resultsall}

\subsection{The non-detection of the optical afterglow} \label{section:OA}

It is quickly seen that \thisgrb\ falls into the category of the
so-called ``dark bursts''. Using, for example, the quick criterion
proposed by \citet{jakobsson2003:apj617:21}, we find $\beta_{OX} <
-0.05$ at 12.7 hours after the burst using the Gemini $r^{\prime}$
band observation, well below the proposed limit of $\beta_{OX} < 0.5$.
A more precise criterion would combine the available spectral and
temporal parameters of the X-ray afterglow, allow all valid
combinations, and from that infer the range of possible optical
magnitudes from the X-rays \citep[see e.g.][]{rol2005:apj624:868}.
This is, in fact implied in our previous modeling: the modeled
specific fluxes corresponding to the band and epoch of our optical and
nIR upper limits are listed in Table \ref{table:compupperlimit} (see
also Table \ref{table:limmags}).

While the values in this table are given for local extinction, not
K-corrected to $z = 0.8$, it is
immediately obvious that our $K$-band observations put a 
stringent constraint on the required extinction directly surrounding
the burst.

\begin{deluxetable}{cccccccc}
  \tablecaption{Upper limits compared to model specific flux calculations.  The
    inferred lower limits on the extinction are given in the observers
    frame. The \ebv\ values are given for a Galactic extinction curve
    ($R_V = 3.08$), and are for illustrative purposes; see the
    comments at the end of Section \ref{section:OA}.
\label{table:compupperlimit}}
\tablehead{
\colhead{filter} &
\colhead{upper limit} &
\multicolumn{3}{c}{homogeneous density profile} &
\multicolumn{3}{c}{stellar wind density profile} \\
&
& 
\colhead{modeled specific flux} &
\colhead{extinction} &
\colhead{\ebv} &
\colhead{modeled specific flux} &
\colhead{extinction} &
\colhead{\ebv} \\
&
\colhead{($\mu$Jy)} &
\colhead{($\mu$Jy)} &
\colhead{(mag.)}&
\colhead{} &
\colhead{($\mu$Jy)} &
\colhead{(mag.)}&
\colhead{}
}

\startdata
$K_s$    &   $< 6.82 $  &  $93.1$  &  2.84  &  7.74   &  $57.2$  &  2.31  &  6.29 \\
$J$      &   $<12.3  $  &  $117 $  &  2.44  &  2.71   &  $74.1$  &  1.95  &  2.16 \\
$Z$      &   $< 2.66 $  &  $103 $  &  3.97  &  2.58   &  $67.8$  &  3.52  &  2.29 \\
$r^{\prime}$     &   $< 0.305$  &  $74.5$  &  5.97  &  2.17   &  $44.4$  &  5.41  &  1.97 \\
$z^{\prime}$     &   $< 1.53 $  &  $87.7$  &  4.40  &  2.97   &  $51.9$  &  3.83  &  2.59 \\

\enddata
\end{deluxetable}

To estimate the amount of local extinction in the host galaxy, we have
modeled the nIR to X-ray spectrum around 0.5~days after the burst,
considering 3 different extinction curves: those of the Milky Way
(MW), the Large Magellanic Cloud (LMC) and the Small Magellanic Cloud
(SMC), from \citet{pei1992:apj395:130}, with $R_V$ of 3.08, 3.16 and
2.93, respectively.

For this, we used the unabsorbed XRT flux obtained from the spectral
fit to orbits 3~--~7 (which do not contain piled-up data), and fixed
the energy spectral slope in the X-rays at $\beta = 1$ (also from the
X-ray spectral fit). The optical specific fluxes were scaled to the
logarithmic mid-observation time of the X-ray observations with an
assumed $\alpha = 1.16$ decline. This estimated optical decay is
derived from the pre-break X-ray decay value, allowing for the cooling
break between the two wavelength regions, and averaging the two
possible values for $\alpha_X - \alpha_{\mathrm{opt}}$ (-0.25 and
0.25).  We can further put the most stringent constraint on the broken
power law spectral shape, by setting the spectral break just below the
X-rays, at $1.8 \times 10^{17}$~Hz, which follows from our previous
broad-band modeling. Our results indicate that, for the aforementioned
extinction curves, a local extinction of $\ebv \approx 7$ (for all
three extinction curves) is necessary to explain the $K$-band upper
limit.

We can relate the resulting \nh\ from our X-ray spectral fits to any
local \ebv, using the relations found in \citet{predehl1995:aa293},
\citet{fitzpatrick1985:apj299} and \citet{martin1989:aa215:219} for
$N(HI)/\ebv$, and adjusting the metallicity in our X-ray absorption
model accordingly. We obtain $\ebv = 7.5, 1.54$ and 0.84 for a MW, LMC
and SMC extinction curve respectively, with the MW value showing the
best agreement with our findings for optical extinction (both
\citealt{nakagawa2006:pasjl58:35} and \citealt{butler2005:gcn4170}
find \ebv\ values roughly twice as high here, for a MW extinction
curve only, since their \nh\ estimate is larger than ours). This,
obviously, depends on the assumption that the MW (or otherwise, LMC or
SMC) extinction curves are valid models to compare with our observed
data here. Since these data happen to originate from just one sight
line in a galaxy, this may very well not be the case.  Further, even
if the extinction curve is correct, the actual value of $R_V$ may be
rather different for the host galaxy. Finally, the \ebv~--~\nh\
relations show a rather large scatter, especially at higher column
densities, nor is the \nh\ always derived using X-ray spectroscopy.
Our above results are therefore approximations, which are useful to
compare with other (GRB host) studies, but should be taken with the
necessary caution.

\subsection{The host galaxy of GRB\,051022}
\label{section:host}

\begin{deluxetable}{ccc}
  \tablecaption{Measured host galaxy magnitudes\label{table:hostgalaxy}}
\tablehead{
\colhead{filter} &
\colhead{magnitude} &
\colhead{magnitude error}
}
\startdata
$K$                   &  18.40  &  0.04  \\ 
$K_s$                 &  18.36  &  0.09  \\
$H$                   &  19.42  &  0.09  \\
$J$                   &  19.92  &  0.05  \\
$Z$\tablenotemark{a}  &  21.41  &  0.05  \\
$z^{\prime}$           &  21.30  &  0.04  \\
$i^{\prime}$           &  21.77  &  0.01  \\
$r^{\prime}$           &  22.04  &  0.01  \\
$R$                   &  21.84  &  0.09  \\
$V$                   &  22.30  &  0.04  \\
$B$                   &  22.75  &  0.02  \\
$U$                   &  $>$ 21.3\tablenotemark{b} & \nodata
\enddata
\tablenotetext{a}{AB magnitude}
\tablenotetext{b}{5-$\sigma$ upper limit}
\end{deluxetable}

Using the optical data described above, we fit the SED of the host of
GRB\,051022 using the HyperZ program\footnote{See
  \url{http://webast.ast.obs-mip.fr/hyperz}} developed by
\citet{bolzonella2000:aap363:476}. The photometry of the host has been
performed using \swcommand{apphot} within IRAF, in an aperture 1.5
times the estimated seeing in the different exposures. The results are
reported in Table \ref{table:hostgalaxy} (see also
\citealt{ovaldsen2007:apj662:294}). The range of photometric
magnitudes reported in this paper provides one of the most complete
broadband optical datasets of a GRB host galaxy to date.  We fit using
the eight synthetic galaxy templates provided within HyperZ at the
redshift of the host, and find that the host galaxy is a blue compact
galaxy of type irregular, with a dominant stellar population age of
$\approx 20$~Myr, similar to other long GRB hosts
\citep{christensen2005:apjl631:29}.  A moderate amount of extinction
of $A_V \approx 1$~mag is required to fit the SED, with an SMC-type
extinction curve providing a best fit, and the luminosity of the host
is approximately 1.5 $L_*$ (assuming $M_{*,B} = -21$); these findings
are in full agreement with \citet{castro-tirado2006:aipc836:79}.  The
amount of extinction in the line of sight towards the GRB required to
suppress the optical light of the afterglow to the observed limits is
clearly higher than the $A_V$ value found from the host SED: $A_V =
4.4$ magnitudes towards the GRB, estimated from blueshifting our
measured (observer frame) $z^{\prime}$ band extinction to $z = 0.8$.
The host galaxy SED extinction is, however, an average value derived
from the integrated colors of the host.

The host of GRB\,051022 is located in a field crowded with galaxies of
various Hubble types.  We perform photometry on several galaxies close
to the GRB host (within 1 arcminute) to investigate the possibility
that the high star formation rate seen in the optical
(\citealt{castro-tirado2006:aipc836:79} report an SFR of
$\approx\,20$M$_\odot$yr$^{-1}$) is induced by a recent interaction
with one of the neighboring galaxies. As formation of high mass stars
has also been observed to occur in dusty regions in merging systems
\citep[see e.g.][]{lin2007:apjl660:51}, this could help to explain the
excess optical extinction towards GRB\,051022. We performed HyperZ
fits to these galaxies, and find that none of them is well fit by a
photometric redshift of $z \approx 0.8$. Particularly the two galaxies
closest to the GRB host galaxy are not compatible with a redshift 0.8,
and show best fits with photometric redshifts of $z \approx
0.2$~--~0.25.  Out of the sample of six galaxies close to the GRB host
we find that four have best-fit photometric redshifts in the range
0.20~--~0.25, making it unlikely that a possible overdensity of
galaxies near the host galaxy is due to a cluster or galaxy group at
the host redshift.

\section{Discussion} \label{section:discussion}

The issue of non-detected (``dark'') GRB afterglows has received
significant interest ever since the discovery of the first GRB
afterglow, starting with the non-detection of GRB\,970828 to very deep
limits \citep{groot1998:apj493:27,odewahn1997:iauc6735}. For this
particular afterglow, its non-detection has been attributed to a
dust-lane in its host galaxy \citep{djorgovski2001:apj562}. Dust
extinction as the cause of the non-detection of the optical afterglow
has been inferred in the case of several other GRBs, notably those
with a precise X-ray or radio position, where one can pinpoint the
afterglow position on top of its host galaxy (e.g. GRB\,000210,
\citealt{piro2002:apj577}).

Optical drop-outs due to high redshift will also result in dark
bursts, but are harder to confirm, since it would require at least one
detection in a red band, to detect the Ly$\alpha$ break. Otherwise, it
becomes indistinguishable from dust extinction.

Other explanations of afterglow non-detections include the intrinsic
faintness of the afterglow. For HETE-2 detected GRBs, this has been
inferred for e.g. GRB\,020819 \citep{jakobsson2005:apj629:45}. For
Swift bursts, where rapid and accurate X-ray positions are often
available, this is a major cause of non-detections
\citep{berger2005:apj634:501}, largely attributed to a higher average
redshift.

In our case here, the host galaxy has been detected at a relatively
modest redshift, which almost automatically points to the dust
extinction scenario. The radio and X-ray detections even allow us to
accurately model the necessary amount of extinction between us and the
GRB.

\subsection{The burst environment} \label{section:environment}

The issue of the role of dust extinction in the lines of sight towards
GRBs is still very much an open one. While clear signs of dust
depletion are seen in several afterglow spectra, the $A_V$ values that
are predicted from these depletion measures are generally much higher
than the observed ones, that can be found from the continuum shape
\citep{savaglio2004:apj614:293}. Recently, selected samples of GRB
afterglows were homogeneously analyzed for X-ray and optical
extinction, showing dust to gas ratios different from Galactic and
Magellanic cloud values
\citep{starling2007:apj661:787,schady2007:mnras377:273}. 
\citet{galama2001:apj549} and \citet{stratta2004:apj608:846} had
already found dust (optical) to gas (X-ray) ratios to be lower than
the Galactic value (in all cases, however, there is a bias in
these samples to optically and X-ray detected afterglows).  Comparison
of neutral hydrogen columns and metallicities of afterglow lines of
sight with X-ray extinction values \citep{watson2007:apjl600:101}
showed that the absorption probed by these two wavelength regimes is
generally located at different positions in the host. In all these
cases there may be significant biases against bursts with low apparent
magnitudes, preventing optical spectroscopy, which are hard to
quantify.

In the case of \thisgrb\, there is a significant discrepancy between
the extinction for the host as a whole and that along the line of
sight to the burst, or at least along our line of sight towards the
burst. This is perhaps not too surprising if one assumes, for example,
that the burst occurred inside a Giant Molecular Cloud (GMC).
\citet{jakobsson2006:aal460:13} compared the GRB $N$(HI) distribution
to that of modeled GRBs located inside Galactic-like GMCs. They found
that the two distributions are incompatible, and possibly GRBs are
more likely to occur inside clouds with a lower $N$(HI), or
alternatively, outside the actual GMC. (Note that their study
concentrates on bursts with $z > 2$, where the Ly-$\alpha$ absorption
is visible in the optical wavebands; it is also biased towards
optically detected afterglows). A GMC could therefore actually be
positioned in front of the GRB, where the required optical and X-ray
extinction is easily achieved. This agrees with the findings by
\citet{prochaska2007:astro-ph/0703365}, who analyzed several
GRB-Damped\,Lyman\,Alpha spectra and from observed depletion levels
infer that the gas is not located directly near the GRB (e.g. its
molecular cloud) but further out. The specific case of GRB\,060418
confirmed this through time-resolved high resolution spectroscopy,
showing that the observed metal lines originate past 1.7 kpc from the
burst itself \citep{vreeswijk2007:aap468:83}. In fact, X-ray radiation
from the burst could easily destroy grains out to 100 pc
\citep{waxman2000:apj537:796,fruchter2001:apj563:597,drain2002:apj569:780}
and permit the afterglow radiation to penetrate the surrounding
molecular cloud. Dust extinction is therefore likely to occur further
out, perhaps to several kiloparsecs.

It is interesting to find a non-SMC type of extinction curve from the
combination of X-ray and optical absorption (though not completely
ruled out): in most cases modeled, an SMC extinction curve fits the
optical--X-ray spectra best
\citep{starling2007:apj661:787,schady2007:mnras377:273}, presumably
attributable to the absence of the 2175 \AA\ feature
\citep{savage1979:araa17:73} and the low dust to gas ratio. Our
findings indicate that the extinction along the line of sight to the
GRB will generally be different than one of the three assumed
extinction curves.  Local small scale density variations in clouds,
such as found by from infrared studies in the Taurus region and from
simulations \citep{padoan2006:apj649:807}, could cause this fairly
easily.

\subsection{Energetics}

Our modeling provides us with a detailed set of parameters of the
afterglow energetics, including \Ejet, the energy of the afterglow.
For the prompt emission energy, we use the data from the Konus-Wind
measurements \citep{golenetskii2005:gcn4150}. We calculate a prompt
isotropic energy of $4.39\plusminus{0.29}{0.18} \times 10^{53}$ erg in
the 20 keV~--~20 MeV observer frame, and, by applying a K-correction
\citep[as in e.g.][]{bloom2001:aj121}, $\Epromptiso =
10.4\plusminus{0.7}{0.4} \times 10^{53}$ erg in the 1~--~$10^5$ keV
rest frame. The collimation corrected energy depends on the assumed
density profile of the surrounding medium: for a homogeneous medium, we
obtain $\Epromptjet = 18.2 \times 10^{50}$ erg, and for a wind-like
medium, $\Epromptjet = 8.38 \times 10^{50}$ erg.  With $\Epeak =
918\plusminus{66}{59}$ keV in the burst rest frame, we find that the
\Epeak~--~\Epromptjet\ relation \citep{ghirlanda2004:apj616:331}
somewhat underestimates the \Epeak\ when calculated from \Epromptjet:
$\Epeak \approx 740$ keV for a homogeneous medium, and $\approx 430$
keV for a wind medium (the difference between our chosen cosmology and
that used by \citealt{ghirlanda2004:apj616:331} amounts to only a
0.3\% difference in \Eiso).  These estimates, however, come with a few
caveats: \emph{1)} the \Epeak\ from the Konus-Wind data is calculated
using an exponential cut-off model, not the Band function
\citep{band1993:apj413:281}. Since the Band function includes the case
of an exponential cut-off model (with $\beta = -\infty$, this should,
however, pose no problem in estimating the actual \Epeak), \emph{2)}
our break time, and therefore the jet-opening angle, are calculated
from the full modeling of the afterglow, which effectively means
derived from the available X-ray and radio data. Further, the original
Ghirlanda relation was derived using optical break times. Recent
efforts show that estimating jet-break times from X-ray light curves
may not lead to the same results
\citep[e.g.][]{panaitescu2006:mnras369:2059}, and \emph{3)} the
relatively large error on the jet opening angle estimate allows for a
relatively large range in collimation corrected energies. We have
simply used here our best value, but an \Epeak\ value of 1498 keV
derived from \Ejet\ can still be accommodated within our errors. (We
note that, with a different \Epeak\ estimate and an incorrect value
for the jet-break time, \citealt{nakagawa2006:pasjl58:35} still found
their results to lie on the Ghirlanda relation).  The break time
problem can be avoided by looking only at the \Epeak\ -- \Epromptiso\
relation \citep{amati2002:aa390:81,amati2006:mnras372:233}. From this,
we estimate $\Epeak \approx 924$ keV, nicely in agreement with the
value found directly from the spectra fit.

Comparing the prompt emission energy (\Epromptjet) and afterglow blast
wave kinetic energy (\Ejet), we find their ratio to be $\Epromptjet /
\Ejet = 3.7$ in the case of a wind-like circumburst medium, while for
a homogeneous medium, $\Epromptjet / \Ejet = 20$.  These ratios are
similar to those found for other bursts \citep[e.g.][Figure
3]{berger2003:nature426:154}.

\thisgrb\ is also one of the brightest bursts observed by HETE, with a
prompt 30--400 keV fluence of $S = 1.31 \times 10^{-4}$ erg cm$^{-2}$
\citep{nakagawa2006:pasjl58:35}. In fact, compared to the sample of 35
FREGATE bursts analyzed by \citet{barraud2003:aa400}, \thisgrb\ has
the largest fluence, even topping the relatively close-by GRB\,030329
(\citealt[$S = 1.2 \times 10^{-4}$ erg
cm$^{-2}$]{vanderspek2004:apj617:1251}; note that for \thisgrb, its
redshift is close to the median redshift of HETE-2 detected GRBs and
therefore distance effects will play a very minor role).
\citet{rol2005:apj624:868} noted this potential correlation of fluence
with the non-detection of a GRB afterglow for the small subset of
genuinely dark bursts in their sample: the truly dark bursts all have
a much higher than average fluence (although this is for a relatively
small sample only).  Potentially, this could point to an external
origin for the prompt emission, instead of being due to internal
shocks: a large amount of dust may result in more matter that will
radiate, while at the same time the radiation will be suppressed at UV
and optical wavelengths.  This would indicate an origin of the
extinction quite close to the burst instead, in contrast to previous
findings for other bursts, as discussed in Section
\ref{section:environment}. These latter bursts, however, were all
optically selected to obtain spectroscopy, and may therefore show
different surroundings than \thisgrb.  Unfortunately, with the small
sample size of genuine dark bursts a firm conclusion on this
correlation is not possible, but remains something to watch for in
future dark bursts.

\section{Conclusions} \label{section:conclusions}

\thisgrb\ is a prototypical dark burst, with the local extinction
exceeding $2.3$ magnitudes in $J$ and 5.4 magnitudes in $U$, in the
host-galaxy restframe, with the exact limits depending on the
circumburst density profile.  The extinction curve derived from an
X-ray~--~optical spectral fit points towards a Galactic type of
extinction curve, although it is likely that this is more or less a
coincidence: the host galaxy itself is best modeled with an SMC-like
extinction curve, with a modest amount of extinction, $A_V \approx
1$~mag.  The large optical absorption towards the afterglow of
GRB\,051022 is therefore probably the effect of an unfortunate
position in the host where the line of sight crosses dense regions
within the host.

The X-ray and radio afterglow data allow for a full solution of the
blastwave model, although we unfortunately cannot distinguish between
the density profile (homogeneous or wind-like) of the circumburst
medium.  We estimate a collimation-corrected energy in the afterglow
emission of 0.92~--~2.3 $\times 10^{50}$ erg, while the energy in
prompt emission (1~--~$10^5$~keV rest frame) is 8.4~--~18 $\times
10^{50}$~erg. Aside from the large optical extinction, the afterglow
otherwise appears as an average afterglow, with no outstanding
properties. The potentially interesting point here is that the
30-400~keV fluence of the prompt emission is one of the largest ever
detected in the HETE-2 sample.

In the era of Swift GRBs, dust-extincted bursts can actually be
found in optical/nIR thanks to the rapid availability of precise
positions: examples are found where the burst is relatively bright
early on at optical/nIR wavelengths, while the afterglow proper (post
few hours) often can go undetected
\citep[e.g.][]{oates2006:mnras372:3270,perley2007:astro-ph/0703538}.
This allows targeted follow-up of such dark bursts, i.e. determining
the host galaxy (and the bursts precise position therein) and a
redshift measurement. In our case, a precise CXO and radio position
pinpointed the host galaxy, but such data may not always be available.
High resolution late-time observations of the host, at the location of
the GRB, may then reveal whether the burst indeed occurred inside a
dense host region.

\acknowledgments

{\small We thank the referee for a careful reading of the manuscript
  and constructive comments.  We thank Kim Page and Andy Beardmore for
  useful discussions regarding the XRT data analysis. ER and RLCS
  acknowledge support from PPARC. KW and RAMJW acknowledge support of
  NWO under grant 639.043.302. The authors acknowledge funding for the
  Swift mission in the UK by STFC, in the USA by NASA and in Italy by
  ASI. The Dark Cosmology Centre is funded by the Danish National
  Research Foundation. The William Herschel Telescope is operated on
  the island of La Palma by the Isaac Newton Group in the Spanish
  Observatorio del Roque de los Muchachos of the Instituto de
  Astrof\'{\i}sica de Canarias.  The United Kingdom Infrared Telescope
  is operated by the Joint Astronomy Centre on behalf of the U.K.
  Particle Physics and Astronomy Research Council. The data reported
  here were obtained as part of the UKIRT Service Programme. The
  Westerbork Synthesis Radio Telescope is operated by ASTRON
  (Netherlands Foundation for Research in Astronomy) with support from
  the Netherlands Foundation for Scientific Research (NWO). Support
  for this work was provided by the National Aeronautics and Space
  Administration through Chandra Award Number 1736937 issued by the
  Chandra X-ray Observatory Center, which is operated by the
  Smithsonian Astrophysical Observatory for and on behalf of the
  National Aeronautics Space Administration under contract NAS8-03060.
  This publication makes use of data products from the Two Micron All
  Sky Survey, which is a joint project of the University of
  Massachusetts and the Infrared Processing and Analysis
  Center/California Institute of Technology, funded by the National
  Aeronautics and Space Administration and the National Science
  Foundation. This research has made use of data obtained from the
  High Energy Astrophysics Science Archive Research Center (HEASARC),
  provided by NASA's Goddard Space Flight Center. }

\appendix

\section{Interstellar scintillation in the radio modeling}

The 4.9 GHz measurements show scatter around the best fit light curve,
which can be accounted for by interstellar scintillation (ISS).  In
Figure \ref{figure:resultsall} we have indicated the predicted rms
scatter due to ISS.  We have calculated the scattering measure from
the \citet{cordes2002:astro-ph/0207156} model for the Galactic
distribution of free electrons: $SM=2.04\cdot
10^{-4}\,\mathrm{kpc}\,/\mathrm{m}^{-20/3}$.  The radio specific flux will be
modulated when the source size is close to one of the three
characteristic angular scales, i.e. for weak, refractive or
diffractive ISS.  From \citet{walker1998:mnras294:307}, we calculate
the transition frequency between weak and strong ISS,
$\nu_0=9.12\,\mathrm{GHz}$, and the angular size of the first Fresnel
zone, $\theta_{{\mathrm{F}}_0}=0.994\,\mu\mathrm{as}$.  Our
measurements were all performed at frequencies below $\nu_0$, i.e. in
the strong ISS regime, which means that only refractive and
diffractive ISS modulate the specific flux significantly.  We calculate the
evolution of the source size in the extreme relativistic phase
($\theta_\mathrm{s}=R/\Gamma$) and after the jet-break
($\theta_\mathrm{s}=R \theta_{\mathrm{j}}$), and compare this source
size with the diffractive angular scale
$\theta_\mathrm{d}=\theta_{{\mathrm{F}}_0}(\nu_0/\nu)^{-6/5}=0.0701\cdot
\nu_{\mathrm{GHz}}^{6/5}\,\mu\mathrm{as}$ and the refractive angular
scale
$\theta_\mathrm{r}=\theta_{{\mathrm{F}}_0}(\nu_0/\nu)^{11/5}=128\cdot
\nu_\mathrm{GHz}^{-11/5}\,\mu\mathrm{as}$ to calculate the modulation
index $m_\mathrm{p}$.  In the case of diffractive ISS the modulation
index is $1$, and in the case of refractive ISS
$m_\mathrm{p}=(\nu_0/\nu)^{-17/30}=0.286\cdot
\nu_{\mathrm{GHz}}^{17/30}$.  Because of the expansion of the
blastwave the angular source size exceeds one of the characteristic
angular scales at some point in time.  Then the modulation will begin
to quench as $m_\mathrm{p}(\theta_\mathrm{d}/\theta_\mathrm{s})$ in
the case of diffractive ISS, and as
$m_\mathrm{p}(\theta_\mathrm{r}/\theta_\mathrm{s})^{7/6}$ in the case
of refractive ISS.

\bibliographystyle{apj}
\bibliography{references}

\begin{thebibliography}{85}
\expandafter\ifx\csname natexlab\endcsname\relax\def\natexlab#1{#1}\fi

\bibitem[{{Alard}(2000)}]{alard2000:aaps144}
{Alard}, C. 2000, \aaps, 144, 363

\bibitem[{{Amati}(2006)}]{amati2006:mnras372:233}
{Amati}, L. 2006, \mnras, 372, 233

\bibitem[{{Amati} {et~al.}(2002){Amati}, {Frontera}, {Tavani}, {in't Zand},
  {Antonelli}, {Costa}, {Feroci}, {Guidorzi}, {Heise}, {Masetti}, {Montanari},
  {Nicastro}, {Palazzi}, {Pian}, {Piro}, \& {Soffitta}}]{amati2002:aa390:81}
{Amati}, L., {Frontera}, F., {Tavani}, M., {et~al.} 2002, \aap, 390, 81

\bibitem[{{Arnaud}(1996)}]{arnaud1996:proc:17}
{Arnaud}, K.~A. 1996, in ASP Conf. Ser. 101: Astronomical Data Analysis
  Software and Systems V, 17

\bibitem[{{Band} {et~al.}(1993){Band}, {Matteson}, {Ford}, {Schaefer},
  {Palmer}, {Teegarden}, {Cline}, {Briggs}, {Paciesas}, {Pendleton}, {Fishman},
  {Kouveliotou}, {Meegan}, {Wilson}, \& {Lestrade}}]{band1993:apj413:281}
{Band}, D., {Matteson}, J., {Ford}, L., {et~al.} 1993, \apj, 413, 281

\bibitem[{{Barraud} {et~al.}(2003){Barraud}, {Olive}, {Lestrade}, {Atteia},
  {Hurley}, {Ricker}, {Lamb}, {Kawai}, {Boer}, {Dezalay}, {Pizzichini},
  {Vanderspek}, {Crew}, {Doty}, {Monnelly}, {Villasenor}, {Butler}, {Levine},
  {Yoshida}, {Shirasaki}, {Sakamoto}, {Tamagawa}, {Torii}, {Matsuoka},
  {Fenimore}, {Galassi}, {Tavenner}, {Donaghy}, {Graziani}, \&
  {Jernigan}}]{barraud2003:aa400}
{Barraud}, C., {Olive}, J.-F., {Lestrade}, J.~P., {et~al.} 2003, \aap, 400,
  1021

\bibitem[{{Berger} {et~al.}(2005){Berger}, {Kulkarni}, {Fox}, {Soderberg},
  {Harrison}, {Nakar}, {Kelson}, {Gladders}, {Mulchaey}, {Oemler}, {Dressler},
  {Cenko}, {Price}, {Schmidt}, {Frail}, {Morrell}, {Gonzalez}, {Krzeminski},
  {Sari}, {Gal-Yam}, {Moon}, {Penprase}, {Jayawardhana}, {Scholz}, {Rich},
  {Peterson}, {Anderson}, {McNaught}, {Minezaki}, {Yoshii}, {Cowie}, \&
  {Pimbblet}}]{berger2005:apj634:501}
{Berger}, E., {Kulkarni}, S.~R., {Fox}, D.~B., {et~al.} 2005, \apj, 634, 501

\bibitem[{{Berger} {et~al.}(2003){Berger}, {Kulkarni}, {Pooley}, {Frail},
  {McIntyre}, {Wark}, {Sari}, {Soderberg}, {Fox}, {Yost}, \&
  {Price}}]{berger2003:nature426:154}
{Berger}, E., {Kulkarni}, S.~R., {Pooley}, G., {et~al.} 2003, \nat, 426, 154

\bibitem[{{Berger} \& {Wyatt}(2005)}]{berger2005:gcn4148}
{Berger}, E. \& {Wyatt}, P. 2005, GCN Circular, 4148

\bibitem[{{Bloom} {et~al.}(2001){Bloom}, {Frail}, \& {Sari}}]{bloom2001:aj121}
{Bloom}, J.~S., {Frail}, D.~A., \& {Sari}, R. 2001, \aj, 121, 2879

\bibitem[{{Bolzonella} {et~al.}(2000){Bolzonella}, {Miralles}, \&
  {Pell{\'o}}}]{bolzonella2000:aap363:476}
{Bolzonella}, M., {Miralles}, J.-M., \& {Pell{\'o}}, R. 2000, \aap, 363, 476

\bibitem[{{Bremer} {et~al.}(2005){Bremer}, {Castro-Tirado}, \&
  {Neri}}]{bremer2005:gcn4157}
{Bremer}, M., {Castro-Tirado}, A.~J., \& {Neri}, R. 2005, GCN Circular, 4157

\bibitem[{{Butler} {et~al.}(2005{\natexlab{a}}){Butler}, {Ricker}, {Lamb},
  {Burrows}, {Racusin}, \& {Gehrels}}]{butler2005:gcn4165}
{Butler}, N.~R., {Ricker}, G.~R., {Lamb}, D.~Q., {et~al.} 2005{\natexlab{a}},
  GCN Circular, 4165

\bibitem[{{Butler} {et~al.}(2005{\natexlab{b}}){Butler}, {Ricker}, {Lamb},
  {Burrows}, {Racusin}, \& {Gehrels}}]{butler2005:gcn4170}
---. 2005{\natexlab{b}}, GCN Circular, 4170

\bibitem[{{Cameron} \& {Frail}(2005)}]{cameron2005:gcn4154}
{Cameron}, P.~B. \& {Frail}, D.~A. 2005, GCN Circular, 4154

\bibitem[{{Castro-Tirado} {et~al.}(2006){Castro-Tirado}, {McBreen},
  {Jel{\'{\i}}nek}, {Pandey}, {Bremer}, {de Ugarte Postigo}, {Gorosabel},
  {Guziy}, {Bihain}, {Caballero}, {Ferrero}, {de Jong}, {Misra}, \&
  {Sahu}}]{castro-tirado2006:aipc836:79}
{Castro-Tirado}, A.~J., {McBreen}, S., {Jel{\'{\i}}nek}, M., {et~al.} 2006, in
  American Institute of Physics Conference Series, Vol. 836, Gamma-Ray Bursts
  in the Swift Era, ed. S.~S. {Holt}, N.~{Gehrels}, \& J.~A. {Nousek}, 79--84

\bibitem[{{Christensen} {et~al.}(2005){Christensen}, {Hjorth}, \&
  {Gorosabel}}]{christensen2005:apjl631:29}
{Christensen}, L., {Hjorth}, J., \& {Gorosabel}, J. 2005, \apjl, 631, L29

\bibitem[{{Cordes} \& {Lazio}(2002)}]{cordes2002:astro-ph/0207156}
{Cordes}, J.~M. \& {Lazio}, T.~J.~W. 2002, astro-ph/0207156

\bibitem[{{Dickey} \& {Lockman}(1990)}]{dickey1990:araa28:215}
{Dickey}, J.~M. \& {Lockman}, F.~J. 1990, \araa, 28, 215

\bibitem[{{Djorgovski} {et~al.}(2001){Djorgovski}, {Frail}, {Kulkarni},
  {Bloom}, {Odewahn}, \& {Diercks}}]{djorgovski2001:apj562}
{Djorgovski}, S.~G., {Frail}, D.~A., {Kulkarni}, S.~R., {et~al.} 2001, \apj,
  562, 654

\bibitem[{{Draine} \& {Hao}(2002)}]{drain2002:apj569:780}
{Draine}, B.~T. \& {Hao}, L. 2002, \apj, 569, 780

\bibitem[{{Evans} {et~al.}(2007){Evans}, {Beardmore}, {Page}, {Tyler},
  {Osborne}, {Goad}, {O'Brien}, {Vetere}, {Racusin}, {Morris}, {Burrows},
  {Capalbi}, {Perri}, {Gehrels}, \& {Romano}}]{evans2007:aap469:379}
{Evans}, P.~A., {Beardmore}, A.~P., {Page}, K.~L., {et~al.} 2007, \aap, 469,
  379

\bibitem[{{Fitzpatrick}(1985)}]{fitzpatrick1985:apj299}
{Fitzpatrick}, E.~L. 1985, \apj, 299, 219

\bibitem[{{Fruchter} {et~al.}(2001){Fruchter}, {Krolik}, \&
  {Rhoads}}]{fruchter2001:apj563:597}
{Fruchter}, A., {Krolik}, J.~H., \& {Rhoads}, J.~E. 2001, \apj, 563, 597

\bibitem[{{Fynbo} {et~al.}(2001){Fynbo}, {Jensen}, {Gorosabel}, {Hjorth},
  {Pedersen}, {M{\o}ller}, {Abbott}, {Castro-Tirado}, {Delgado}, {Greiner},
  {Henden}, {Magazz{\` u}}, {Masetti}, {Merlino}, {Masegosa}, {{\O}stensen},
  {Palazzi}, {Pian}, {Schwarz}, {Cline}, {Guidorzi}, {Goldsten}, {Hurley},
  {Mazets}, {McClanahan}, {Montanari}, {Starr}, \&
  {Trombka}}]{fynbo2001:aa369:373}
{Fynbo}, J.~U., {Jensen}, B.~L., {Gorosabel}, J., {et~al.} 2001, \aap, 369, 373

\bibitem[{{Gal-Yam} {et~al.}(2005){Gal-Yam}, {Berger}, {Fox}, {Soderberg},
  {Cenko}, {Cameron}, \& {Frail}}]{gal-yam2005:gcn4156}
{Gal-Yam}, A., {Berger}, E., {Fox}, D.~B., {et~al.} 2005, GCN Circular, 4156

\bibitem[{{Galama} \& {Wijers}(2001)}]{galama2001:apj549}
{Galama}, T.~J. \& {Wijers}, R.~A.~M.~J. 2001, \apjl, 549, L209

\bibitem[{{Ghirlanda} {et~al.}(2004){Ghirlanda}, {Ghisellini}, \&
  {Lazzati}}]{ghirlanda2004:apj616:331}
{Ghirlanda}, G., {Ghisellini}, G., \& {Lazzati}, D. 2004, \apj, 616, 331

\bibitem[{{Golenetskii} {et~al.}(2005){Golenetskii}, {Aptekar}, {Mazets},
  {Pal'shin}, {Frederiks}, \& {Cline}}]{golenetskii2005:gcn4150}
{Golenetskii}, S., {Aptekar}, R., {Mazets}, E., {et~al.} 2005, GCN Circular,
  4150

\bibitem[{{Groot} {et~al.}(1998){Groot}, {Galama}, {Van Paradijs},
  {Kouveliotou}, {Wijers}, {Bloom}, {Tanvir}, {Vanderspek}, {Greiner},
  {Castro-Tirado}, {Gorosabel}, {von Hippel}, {Lehnert}, {Kuijken}, {Hoekstra},
  {Metcalfe}, {Howk}, {Conselice}, {Telting}, {Rutten}, {Rhoads}, {Cole},
  {Pisano}, {Naber}, \& {Schwarz}}]{groot1998:apj493:27}
{Groot}, P.~J., {Galama}, T.~J., {Van Paradijs}, J., {et~al.} 1998, \apjl, 493,
  L27

\bibitem[{{Gunn} \& {Peterson}(1965)}]{gunn1965:apj142:1633}
{Gunn}, J.~E. \& {Peterson}, B.~A. 1965, \apj, 142, 1633

\bibitem[{{Haislip} {et~al.}(2006){Haislip}, {Nysewander}, {Reichart}, {Levan},
  {Tanvir}, {Cenko}, {Fox}, {Price}, {Castro-Tirado}, {Gorosabel}, {Evans},
  {Figueredo}, {MacLeod}, {Kirschbrown}, {Jelinek}, {Guziy}, {Postigo},
  {Cypriano}, {Lacluyze}, {Graham}, {Priddey}, {Chapman}, {Rhoads}, {Fruchter},
  {Lamb}, {Kouveliotou}, {Wijers}, {Bayliss}, {Schmidt}, {Soderberg},
  {Kulkarni}, {Harrison}, {Moon}, {Gal-Yam}, {Kasliwal}, {Hudec}, {Vitek},
  {Kubanek}, {Crain}, {Foster}, {Clemens}, {Bartelme}, {Canterna}, {Hartmann},
  {Henden}, {Klose}, {Park}, {Williams}, {Rol}, {O'Brien}, {Bersier}, {Prada},
  {Pizarro}, {Maturana}, {Ugarte}, {Alvarez}, {Fernandez}, {Jarvis}, {Moles},
  {Alfaro}, {Ivarsen}, {Kumar}, {Mack}, {Zdarowicz}, {Gehrels}, {Barthelmy}, \&
  {Burrows}}]{haislip2006:nat440:181}
{Haislip}, J.~B., {Nysewander}, M.~C., {Reichart}, D.~E., {et~al.} 2006, \nat,
  440, 181

\bibitem[{{Henden}(2005)}]{henden2005:gcn4184}
{Henden}, A. 2005, \gcn, 4184, 1

\bibitem[{{Jakobsson} {et~al.}(2005){Jakobsson}, {Frail}, {Fox}, {Moon},
  {Price}, {Kulkarni}, {Fynbo}, {Hjorth}, {Berger}, {McNaught}, \&
  {Dahle}}]{jakobsson2005:apj629:45}
{Jakobsson}, P., {Frail}, D.~A., {Fox}, D.~B., {et~al.} 2005, \apj, 629, 45

\bibitem[{{Jakobsson} {et~al.}(2006{\natexlab{a}}){Jakobsson}, {Fynbo},
  {Ledoux}, {Vreeswijk}, {Kann}, {Hjorth}, {Priddey}, {Tanvir}, {Reichart},
  {Gorosabel}, {Klose}, {Watson}, {Sollerman}, {Fruchter}, {de Ugarte Postigo},
  {Wiersema}, {Bj{\"o}rnsson}, {Chapman}, {Th{\"o}ne}, {Pedersen}, \&
  {Jensen}}]{jakobsson2006:aal460:13}
{Jakobsson}, P., {Fynbo}, J.~P.~U., {Ledoux}, C., {et~al.} 2006{\natexlab{a}},
  \aap, 460, L13

\bibitem[{{Jakobsson} {et~al.}(2004){Jakobsson}, {Hjorth}, {Fynbo}, {Watson},
  {Pedersen}, {Bj{\"o}rnsson}, \& {Gorosabel}}]{jakobsson2003:apj617:21}
{Jakobsson}, P., {Hjorth}, J., {Fynbo}, J.~P.~U., {et~al.} 2004, \apjl, 617,
  L21

\bibitem[{{Jakobsson} {et~al.}(2006{\natexlab{b}}){Jakobsson}, {Levan},
  {Fynbo}, {Priddey}, {Hjorth}, {Tanvir}, {Watson}, {Jensen}, {Sollerman},
  {Natarajan}, {Gorosabel}, {Castro Cer{\'o}n}, {Pedersen}, {Pursimo},
  {{\'A}rnad{\'o}ttir}, {Castro-Tirado}, {Davis}, {Deeg}, {Fiuza},
  {Mykolaitis}, \& {Sousa}}]{jakobsson2006:aa447:897}
{Jakobsson}, P., {Levan}, A., {Fynbo}, J.~P.~U., {et~al.} 2006{\natexlab{b}},
  \aap, 447, 897

\bibitem[{{Jester} {et~al.}(2005){Jester}, {Schneider}, {Richards}, {Green},
  {Schmidt}, {Hall}, {Strauss}, {Vanden Berk}, {Stoughton}, {Gunn},
  {Brinkmann}, {Kent}, {Smith}, {Tucker}, \& {Yanny}}]{jester2005:aj130:873}
{Jester}, S., {Schneider}, D.~P., {Richards}, G.~T., {et~al.} 2005, \aj, 130,
  873

\bibitem[{{Levan} {et~al.}(2006){Levan}, {Fruchter}, {Rhoads}, {Mobasher},
  {Tanvir}, {Gorosabel}, {Rol}, {Kouveliotou}, {Dell'Antonio}, {Merrill},
  {Bergeron}, {Castro Cer{\'o}n}, {Masetti}, {Vreeswijk}, {Antonelli},
  {Bersier}, {Castro-Tirado}, {Fynbo}, {Garnavich}, {Holland}, {Hjorth},
  {Nugent}, {Pian}, {Smette}, {Thomsen}, {Thorsett}, \&
  {Wijers}}]{levan2006:apj647:471}
{Levan}, A., {Fruchter}, A., {Rhoads}, J., {et~al.} 2006, \apj, 647, 471

\bibitem[{{Lin} {et~al.}(2007){Lin}, {Koo}, {Weiner}, {Chiueh}, {Coil}, {Lotz},
  {Conselice}, {Willner}, {Smith}, {Guhathakurta}, {Huang}, {Le Floc'h},
  {Noeske}, {Willmer}, {Cooper}, \& {Phillips}}]{lin2007:apjl660:51}
{Lin}, L., {Koo}, D.~C., {Weiner}, B.~J., {et~al.} 2007, \apjl, 660, L51

\bibitem[{{Martin} {et~al.}(1989){Martin}, {Maurice}, \&
  {Lequeux}}]{martin1989:aa215:219}
{Martin}, N., {Maurice}, E., \& {Lequeux}, J. 1989, \aap, 215, 219

\bibitem[{{Nakagawa} {et~al.}(2006){Nakagawa}, {Yoshida}, {Sugita}, {Tanaka},
  {Ishikawa}, {Tamagawa}, {Suzuki}, {Shirasaki}, {Kawai}, {Matsuoka}, {Atteia},
  {Pelangeon}, {Vanderspek}, {Crew}, {Villasenor}, {Butler}, {Doty}, {Ricker},
  {Pizzichini}, {Donaghy}, {Lamb}, {Graziani}, {Sato}, {Maetou}, {Arimoto},
  {Kotoku}, {Jernigan}, {Sakamoto}, {Olive}, {Boer}, {Fenimore}, {Galassi},
  {Woosley}, {Yamauchi}, {Takagishi}, \& {Hatsukade}}]{nakagawa2006:pasjl58:35}
{Nakagawa}, Y.~E., {Yoshida}, A., {Sugita}, S., {et~al.} 2006, \pasj, 58, L35

\bibitem[{{Oates} {et~al.}(2006){Oates}, {Mundell}, {Piranomonte}, {Page}, {de
  Pasquale}, {Monfardini}, {Melandri}, {Zane}, {Guidorzi}, {Malesani},
  {Gomboc}, {Bannister}, {Blustin}, {Capalbi}, {Carter}, {D'Avanzo},
  {Kobayashi}, {Krimm}, {O'Brien}, {Page}, {Smith}, {Steele}, \&
  {Tanvir}}]{oates2006:mnras372:3270}
{Oates}, S.~R., {Mundell}, C.~G., {Piranomonte}, S., {et~al.} 2006, \mnras,
  372, 327

\bibitem[{{Odewahn} {et~al.}(1997){Odewahn}, {Djorgovski}, {Kulkarni}, {Frail},
  {Herter}, {Fang}, {Xu}, {Pevunova}, {Steidel}, {Adelberger}, {Kellog},
  {Stanek}, {Garcia}, \& {Krockenberger}}]{odewahn1997:iauc6735}
{Odewahn}, S.~C., {Djorgovski}, S.~G., {Kulkarni}, S.~R., {et~al.} 1997,
  \iaucirc, 6735, 1

\bibitem[{{Oke} \& {Gunn}(1983)}]{oke1983:apj226:713}
{Oke}, J.~B. \& {Gunn}, J.~E. 1983, \apj, 266, 713

\bibitem[{{Olive} {et~al.}(2005){Olive}, {Ricker}, {Atteia}, {Kawai}, {Lamb},
  {Woosley}, {Arimoto}, {Donaghy}, {Fenimore}, {Galassi}, {Graziani},
  {Ishikawa}, {Kobayashi}, {Kotoku}, {Maetou}, {Matsuoka}, {Nakagaw},
  {Sakamoto}, {Sato}, {Shimokawabe}, {Shirasaki}, {Sugita}, {Suzuki},
  {Tamagaw}, {Tanaka}, {Yoshida}, {Butler}, {Crew}, {Doty}, {Prigozhin},
  {Vanderspek}, {Villasenor}, {Jernigan}, {Levine}, {Azzibrouck}, {Braga},
  {Machanda}, {Pizzichini}, {Boer}, {Dezalay}, \& {Hurley}}]{olive2005:gcn4131}
{Olive}, J., {Ricker}, G., {Atteia}, J., {et~al.} 2005, GCN Circular, 4131

\bibitem[{{Ovaldsen} {et~al.}(2007){Ovaldsen}, {Jaunsen}, {Fynbo}, {Hjorth},
  {Thoene}, {Feron}, {Xu}, {Selj}, \& {Teuber}}]{ovaldsen2007:apj662:294}
{Ovaldsen}, J.~., {Jaunsen}, A.~O., {Fynbo}, J.~P.~U., {et~al.} 2007, \apj,
  662, 294

\bibitem[{{Padoan} {et~al.}(2006){Padoan}, {Cambr{\'e}sy}, {Juvela}, {Kritsuk},
  {Langer}, \& {Norman}}]{padoan2006:apj649:807}
{Padoan}, P., {Cambr{\'e}sy}, L., {Juvela}, M., {et~al.} 2006, \apj, 649, 807

\bibitem[{{Panaitescu} \& {Kumar}(2001)}]{panaitescu2001:apj560:49}
{Panaitescu}, A. \& {Kumar}, P. 2001, \apjl, 560, L49

\bibitem[{{Panaitescu} {et~al.}(2006){Panaitescu}, {M{\'e}sz{\'a}ros},
  {Burrows}, {Nousek}, {Gehrels}, {O'Brien}, \&
  {Willingale}}]{panaitescu2006:mnras369:2059}
{Panaitescu}, A., {M{\'e}sz{\'a}ros}, P., {Burrows}, D., {et~al.} 2006, \mnras,
  369, 2059

\bibitem[{{Patel} {et~al.}(2005){Patel}, {Kouveliotou}, \&
  {Rol}}]{patel2005:gcn4163}
{Patel}, S., {Kouveliotou}, C., \& {Rol}, E. 2005, GCN Circular, 4163

\bibitem[{{Pe'er} \& {Wijers}(2006)}]{peer2006:apj643:1036}
{Pe'er}, A. \& {Wijers}, R.~A.~M.~J. 2006, \apj, 643, 1036

\bibitem[{{Pei}(1992)}]{pei1992:apj395:130}
{Pei}, Y.~C. 1992, \apj, 395, 130

\bibitem[{{Perley} {et~al.}(2007){Perley}, {Bloom}, {Butler}, {Pollack},
  {Holtzmann}, {Blake}, {Kocevski}, {Vestrand}, {Li}, {Foley}, {Bellm}, {Chen},
  {Prochaska}, {Starr}, {Filippenko}, {Falco}, {Szentgyorgyi}, {Wren},
  {Wozniak}, {White}, \& {Pergande}}]{perley2007:astro-ph/0703538}
{Perley}, D.~A., {Bloom}, J.~S., {Butler}, N.~R., {et~al.} 2007,
  astro-ph/0703538

\bibitem[{{Piro} {et~al.}(2002){Piro}, {Frail}, {Gorosabel}, {Garmire},
  {Soffitta}, {Amati}, {Andersen}, {Antonelli}, {Berger}, {Frontera}, {Fynbo},
  {Gandolfi}, {Garcia}, {Hjorth}, {in~'t~Zand}, {Jensen}, {Masetti},
  {M{\o}ller}, {Pedersen}, {Pian}, \& {Wieringa}}]{piro2002:apj577}
{Piro}, L., {Frail}, D.~A., {Gorosabel}, J., {et~al.} 2002, \apj, 577, 680

\bibitem[{{Predehl} \& {Schmitt}(1995)}]{predehl1995:aa293}
{Predehl}, P. \& {Schmitt}, J.~H.~M.~M. 1995, \aap, 293, 889

\bibitem[{{Prochaska} {et~al.}(2007){Prochaska}, {Chen}, {Dessauges-Zavadsky},
  \& {Bloom}}]{prochaska2007:astro-ph/0703365}
{Prochaska}, J.~X., {Chen}, H.-W., {Dessauges-Zavadsky}, M., \& {Bloom}, J.~S.
  2007, \apj, submitted

\bibitem[{{Racusin} {et~al.}(2005{\natexlab{a}}){Racusin}, {Burrows}, \&
  {Gehrels}}]{racusin2005:gcn4141}
{Racusin}, J., {Burrows}, D., \& {Gehrels}, N. 2005{\natexlab{a}}, GCN
  Circular, 4141

\bibitem[{{Racusin} {et~al.}(2005{\natexlab{b}}){Racusin}, {Kennea}, {Fox},
  {Burrows}, {Cucchiara}, {Schady}, {Wells}, {Gehrels}, {Sakamoto},
  {Kouveliotou}, \& {Patel}}]{racusin2005:gcn4169}
{Racusin}, J., {Kennea}, J., {Fox}, D., {et~al.} 2005{\natexlab{b}}, GCN
  Circular, 4169

\bibitem[{{Rhoads}(1997)}]{rhoads1997:apj487:487}
{Rhoads}, J.~E. 1997, \apjl, 487, L1

\bibitem[{{Ricker} {et~al.}(2003){Ricker}, {Atteia}, {Crew}, {Doty},
  {Fenimore}, {Galassi}, {Graziani}, {Hurley}, {Jernigan}, {Kawai}, {Lamb},
  {Matsuoka}, {Pizzichini}, {Shirasaki}, {Tamagawa}, {Vanderspek}, {Vedrenne},
  {Villasenor}, {Woosley}, \& {Yoshida}}]{ricker2003:aipc662:3}
{Ricker}, G.~R., {Atteia}, J.-L., {Crew}, G.~B., {et~al.} 2003, in American
  Institute of Physics Conference Series, Vol. 662, Gamma-Ray Burst and
  Afterglow Astronomy 2001: A Workshop Celebrating the First Year of the HETE
  Mission, ed. G.~R. {Ricker} \& R.~K. {Vanderspek}, 3--16

\bibitem[{{Rol} {et~al.}(2007){Rol}, {Osborne}, {Page}, {McGowan}, {Beardmore},
  {O'Brien}, {Levan}, {Bersier}, {Guidorzi}, {Marshall}, {Fruchter}, {Tanvir},
  {Monfardini}, {Gomboc}, {Barthelmy}, \& {Bannister}}]{rol2007:mnras374:1078}
{Rol}, E., {Osborne}, J.~P., {Page}, K.~L., {et~al.} 2007, \mnras, 374, 1078

\bibitem[{{Rol} {et~al.}(2005){Rol}, {Wijers}, {Kouveliotou}, {Kaper}, \&
  {Kaneko}}]{rol2005:apj624:868}
{Rol}, E., {Wijers}, R.~A.~M.~J., {Kouveliotou}, C., {Kaper}, L., \& {Kaneko},
  Y. 2005, \apj, 624, 868

\bibitem[{{Savage} \& {Mathis}(1979)}]{savage1979:araa17:73}
{Savage}, B.~D. \& {Mathis}, J.~S. 1979, \araa, 17, 73

\bibitem[{{Savaglio} \& {Fall}(2004)}]{savaglio2004:apj614:293}
{Savaglio}, S. \& {Fall}, S.~M. 2004, \apj, 614, 293

\bibitem[{{Schady} {et~al.}(2007){Schady}, {Mason}, {Page}, {de Pasquale},
  {Morris}, {Romano}, {Roming}, {Immler}, \& {vanden
  Berk}}]{schady2007:mnras377:273}
{Schady}, P., {Mason}, K.~O., {Page}, M.~J., {et~al.} 2007, \mnras, 377, 273

\bibitem[{{Schaefer}(2005)}]{schaefer2005:gcn4132}
{Schaefer}, B.~E. 2005, GCN Circular, 4132

\bibitem[{{Schlegel} {et~al.}(1998){Schlegel}, {Finkbeiner}, \&
  {Davis}}]{schlegel1998:apj500:525}
{Schlegel}, D.~J., {Finkbeiner}, D.~P., \& {Davis}, M. 1998, \apj, 500, 525

\bibitem[{{Skrutskie} {et~al.}(2006){Skrutskie}, {Cutri}, {Stiening},
  {Weinberg}, {Schneider}, {Carpenter}, {Beichman}, {Capps}, {Chester},
  {Elias}, {Huchra}, {Liebert}, {Lonsdale}, {Monet}, {Price}, {Seitzer},
  {Jarrett}, {Kirkpatrick}, {Gizis}, {Howard}, {Evans}, {Fowler}, {Fullmer},
  {Hurt}, {Light}, {Kopan}, {Marsh}, {McCallon}, {Tam}, {Van Dyk}, \&
  {Wheelock}}]{skrutskie2006:aj131:1163}
{Skrutskie}, M.~F., {Cutri}, R.~M., {Stiening}, R., {et~al.} 2006, \aj, 131,
  1163

\bibitem[{{Starling} {et~al.}(2007){Starling}, {Wijers}, {Wiersema}, {Rol},
  {Curran}, {Kouveliotou}, {van der Horst}, \&
  {Heemskerk}}]{starling2007:apj661:787}
{Starling}, R.~L.~C., {Wijers}, R.~A.~M.~J., {Wiersema}, K., {et~al.} 2007,
  \apj, 661, 787

\bibitem[{{Stratta} {et~al.}(2004){Stratta}, {Fiore}, {Antonelli}, {Piro}, \&
  {De Pasquale}}]{stratta2004:apj608:846}
{Stratta}, G., {Fiore}, F., {Antonelli}, L.~A., {Piro}, L., \& {De Pasquale},
  M. 2004, \apj, 608, 846

\bibitem[{{Tan}(1991)}]{tan1991:aspc19:42}
{Tan}, G.~H. 1991, in Astronomical Society of the Pacific Conference Series,
  Vol.~19, IAU Colloq. 131: Radio Interferometry. Theory, Techniques, and
  Applications, ed. T.~J. {Cornwell} \& R.~A. {Perley}, 42--46

\bibitem[{{Tanvir} {et~al.}(2004){Tanvir}, {Barnard}, {Blain}, {Fruchter},
  {Kouveliotou}, {Natarajan}, {Ramirez-Ruiz}, {Rol}, {Smith}, {Tilanus}, \&
  {Wijers}}]{mnras2004:352:1073}
{Tanvir}, N.~R., {Barnard}, V.~E., {Blain}, A.~W., {et~al.} 2004, \mnras, 352,
  1073

\bibitem[{{Tody}(1986)}]{tody1986:proc:733}
{Tody}, D. 1986, in Instrumentation in astronomy VI; Proceedings of the
  Meeting, Tucson, AZ, Mar. 4-8, 1986. Part 2 (A87-36376 15-35). Bellingham,
  WA, Society of Photo-Optical Instrumentation Engineers, 1986, p. 733., 733

\bibitem[{{Tokunaga} \& {Vacca}(2005)}]{tokunaga2005:pasp117:421}
{Tokunaga}, A.~T. \& {Vacca}, W.~D. 2005, \pasp, 117, 421

\bibitem[{{Torii}(2005)}]{torii2005:gcn4130}
{Torii}, K. 2005, GCN Circular, 4130

\bibitem[{{Trentham} {et~al.}(2002){Trentham}, {Ramirez-Ruiz}, \&
  {Blain}}]{trentham2002:mnras334:983}
{Trentham}, N., {Ramirez-Ruiz}, E., \& {Blain}, A.~W. 2002, \mnras, 334, 983

\bibitem[{{van der Horst} {et~al.}(2007){van der Horst}, {Wijers}, \& {van der
  Horn}}]{vanderhorst2007:submitted}
{van der Horst}, A., {Wijers}, R.~A.~M.~J., \& {van der Horn}, L. 2007, \aap,
  submitted

\bibitem[{{Van der Horst} {et~al.}(2005){Van der Horst}, {Rol}, \&
  {Wijers}}]{vanderhorst2005:gcn4158}
{Van der Horst}, A.~J., {Rol}, E., \& {Wijers}, R.~A.~M.~J. 2005, GCN Circular,
  4158

\bibitem[{{Vanderspek} {et~al.}(2004){Vanderspek}, {Sakamoto}, {Barraud},
  {Tamagawa}, {Graziani}, {Suzuki}, {Shirasaki}, {Prigozhin}, {Villasenor},
  {Jernigan}, {Crew}, {Atteia}, {Hurley}, {Kawai}, {Lamb}, {Ricker}, {Woosley},
  {Butler}, {Doty}, {Dullighan}, {Donaghy}, {Fenimore}, {Galassi}, {Matsuoka},
  {Takagishi}, {Torii}, {Yoshida}, {Boer}, {Dezalay}, {Olive}, {Braga},
  {Manchanda}, \& {Pizzichini}}]{vanderspek2004:apj617:1251}
{Vanderspek}, R., {Sakamoto}, T., {Barraud}, C., {et~al.} 2004, \apj, 617, 1251

\bibitem[{{Vreeswijk} {et~al.}(2007){Vreeswijk}, {Ledoux}, {Smette}, {Ellison},
  {Jaunsen}, {Andersen}, {Fruchter}, {Fynbo}, {Hjorth}, {Kaufer}, {M{\o}ller},
  {Petitjean}, {Savaglio}, \& {Wijers}}]{vreeswijk2007:aap468:83}
{Vreeswijk}, P.~M., {Ledoux}, C., {Smette}, A., {et~al.} 2007, \aap, 468, 83

\bibitem[{{Walker}(1998)}]{walker1998:mnras294:307}
{Walker}, M.~A. 1998, \mnras, 294, 307

\bibitem[{{Watson} {et~al.}(2007){Watson}, {Hjorth}, {Fynbo}, {Jakobsson},
  {Foley}, {Sollerman}, \& {Wijers}}]{watson2007:apjl600:101}
{Watson}, D., {Hjorth}, J., {Fynbo}, J.~P.~U., {et~al.} 2007, \apjl, 660, L101

\bibitem[{{Waxman} \& {Draine}(2000)}]{waxman2000:apj537:796}
{Waxman}, E. \& {Draine}, B.~T. 2000, \apj, 537, 796

\bibitem[{{Yost} {et~al.}(2003){Yost}, {Harrison}, {Sari}, \&
  {Frail}}]{yost2003:apj597:459}
{Yost}, S.~A., {Harrison}, F.~A., {Sari}, R., \& {Frail}, D.~A. 2003, \apj,
  597, 459

\end{thebibliography}

\end{document}